# Overcoming water diffusion limitations in hydrogels via microtubular graphene networks for soft actuators

Margarethe Hauck[1,*], Lena M. Saure[1,*], Berit Zeller-Plumhoff[2,3], Sören Kaps[1], Jörg Hammel[4], Caprice Mohr[1], Lena Rieck[2], Ali Shaygan Nia[5,6], Xinliang Feng[5,6], Nicola M. Pugno[7,8], Rainer Adelung[1,3], Fabian Schütt[1,3,+]

[1]*Functional Nanomaterials, Department of Materials Science, Kiel University, 24143 Kiel, Germany*

[2]*Institute of Metallic Biomaterials, Helmholtz-Zentrum Hereon, Max-Planck-Str. 1, 21502 Geesthacht, Germany*

[3]*Kiel Nano, Surface and Interface Science KiNSIS, Kiel University, 24118 Kiel, Germany*

[4]*Institute of Materials Physics, Helmholtz-Zentrum Hereon, Max-Planck-Str. 1, 21502 Geesthacht, Germany*

[5]*Department of Chemistry and Food Chemistry, Center for Advancing Electronics Dresden (cfaed), Technische Universität Dresden, 01062 Dresden, Germany*

[6]*Max Planck Institute of Microstructure Physics, Weinberg 2, 06120 Halle, Germany*

[7]*Laboratory for Bioinspired, Bionic, Nano, Meta Materials & Mechanics, Department of Civil, Environmental and Mechanical Engineering, University of Trento, via Mesiano 77, I-38123, Trento, Italy*

[8]*School of Engineering and Materials Science, Queen Mary University of London, Mile End Road, London, E1 4NS, UK*

[+]*corresponding author: fas@tf.uni-kiel.de*

*contributed equally*

**Abstract**

Hydrogel-based soft actuators can operate in sensitive environments, bridging the gap of rigid machines interacting with soft matter. However, while stimuli-responsive hydrogels can undergo extreme reversible volume changes of up to ~90%, water transport in hydrogel actuators is in general limited by their poroelastic behavior. For poly(N-isopropylacrylamide) (PNIPAM) the actuation performance is even further compromised by the formation of a dense skin layer. Here we show, that incorporating a bioinspired microtube graphene network into a PNIPAM matrix with a total porosity of only 5.4 % dramatically enhances actuation dynamics by up to ~400 % and actuation stress by ~4000 % without sacrificing the mechanical stability, overcoming the water transport limitations. The graphene network provides both untethered light-controlled and electrically-powered actuation. We anticipate that the concept provides a versatile platform for enhancing the functionality of soft matter by combining responsive and two-dimensional materials, paving the way towards designing soft intelligent matter.

# 1 Introduction

The field of soft robotics aims to overcome the limitations and drawbacks of rigid and hard machines by fabrication of all necessary components, especially actuators, from soft materials, thereby making human-machine interactions and interfaces safer[1,2]. In contrast to stiff parts, soft actuators are able to perform more divers and complex movements and exhibit greater adaptability[3–5]. Prominent materials for soft actuators are smart hydrogels that respond to a stimulus (temperature[6], light[7], pH[8], magnetic[9] and electric fields[10], which can lead for example to a deformation of the material. Above all, for application as soft actuator with appropriate work output and power densities[11,12] these hydrogels need to provide adjustable and controllable response times, deformations, actuation stresses, mechanical properties and cyclic stability. As hydrogels are inherently poroelastic[13], large volumetric deformations over short timescales are limited by the poroelastic time scale of water diffusion[14]. One of the most studied smart hydrogels is the thermo-responsive hydrogel poly(N-isopropylacrylamide) (PNIPAM)[6,15,16] with a lower critical solution temperature (LCST) of about 32 °C[17]. Above this temperature the hydrogel undergoes a reversible phase transition releasing the contained water[18,19]. The deswelling results in a volume change that can be used for actuation. However, due to slow response rates (e.g. 12 % deswelling within 10 min[20]), limiting work output and cycle rate, the properties of bulk PNIPAM hydrogels are not meeting the high requirements needed for soft actuators. In addition to the poroelastic limit, for PNIPAM these limitations are related to the formation of a dense 'skin layer'[21] during the deswelling process, which greatly reduces the water diffusion out of the hydrogel, requiring weeks to reach equilibrium deswelling[22]. Efforts have been made to improve these drawbacks, for example by adding porosity to the hydrogel matrix for an enhanced water transport through the polymer network[14,15,23]. However, an increase in porosity to overcome both the poroelastic limit and skin layer effect has a detrimental effect on the mechanical stability of the hydrogel.[14,24] In order to combine enhanced water diffusion and enhanced Young's modulus, Alsaid et al. have demonstrated that a highly open porous and interconnected network structure can solve this challenge and achieve remarkable deswelling and swelling rates[25]. However, for actuator applications, high actuation stresses and cyclic stability are also required, which are compromised by the introduction of high porosity. Besides the improvement of deswelling and swelling kinetics, providing precise control over the deformation is crucial for the application as soft hydrogel actuator. To achieve accurate actuation, it is essential to have an internal and controllable temperature increase within the hydrogel matrix itself. Photothermal conversion based on light-absorbing filler materials (e.g. carbon-based nanomaterials[26,27], gold nanoparticles[28,29] or polymers[30,31]) allows untethered[32,33] and more flexible operation, but is limited by the area of illumination and depth of light penetration. In contrast, the Joule effect, induced for example by the addition of conductive nanomaterials[34], electrode meshes[35] or wires[36] to the hydrogel, allows volumetric heating of the hydrogel, but usually compromises its softness.

In this work, we present a bioinspired micro- and nanoengineering concept for hydrogel-based actuators by transferring the ubiquitous motif of hierarchical microtubular networks in nature for efficient fluid transport to hydrogels. Compared to the bulk hydrogel, these structured hydrogels, (1) provide improved actuation dynamics (up to + ~400 %) (2) perform increased actuation stress (+ ~4000 %), and (3) provide options for controlled actuation by electrical or photothermal heating, all, while maintaining the mechanical stability and the inherently global soft properties of the hydrogel itself. A network of interconnected hollow graphene microtubes with a total porosity of only 5.4 % and a very low graphene content of 0.35 vol% is incorporated into a PNIPAM hydrogel matrix. This enhances water transport out of and into the hydrogel matrix for improved actuation dynamics without sacrificing mechanical and cyclic stability, thereby overcoming the general limitation of poroelasticity in hydrogels, as well as PNIPAM specific limitations of skin layer formation associated with bulk PNIPAM hydrogels. For both effects we propose a model for the underlying mechanisms. In addition, the approach allows precise control over the composition of the material system, tailoring properties for different applications, such as untethered, light-driven soft actuators and electrically-triggered grippers with control over response times and actuation forces.

**2 Results**

The fabrication method for conductive composite structures of poly(N-isopropylacrylamide) (PNIPAM) and exfoliated graphene (EG) is shown in **Figure S1** and was adapted from a previously reported method[37] for the fabrication of conductive hydrogels. In short, a sacrificial porous network of interconnected tetrapodal zinc oxide (ZnO) (see **Figure S2**) is coated with a thin layer of EG by a wet-chemical drop infiltration process, allowing precise control over the EG concentration within the network. Filling of the free volume with PNIPAM and subsequent removal of ZnO by mild wet-chemical etching results in a PNIPAM hydrogel matrix pervaded by interconnected hollow graphene coated microtubes (see **Figure 1**a). In the following, this will be referred to as PNIPAM-EG (xx vol%), (xx vol% indicates the graphene concentration) whereas pure, but microstructured PNIPAM, containing a network of hollow microtubes but no EG, will be named PNIPAM-structured and the bulk PNIPAM will be entitled PNIPAM-bulk. By adjusting the density of the sacrificial ZnO template, the overall porosity of the microstructured PNIPAM can be tailored. Here, all samples have a porosity of 5.4 %. The 3D rendering of segmented pores from micro computed tomography (microCT) within a region of interest (ROI) of PNIPAM-structured in **Figure 1**b reveals the interconnected network structure of microstructured PNIPAM hydrogels, displaying the microtube network with an interconnectivity of 70 %. The fabrication method enables the preparation of hydrogel-graphene composites of different shapes and geometries as well as with variable graphene content and patterns (see **Figure 1**c and d).

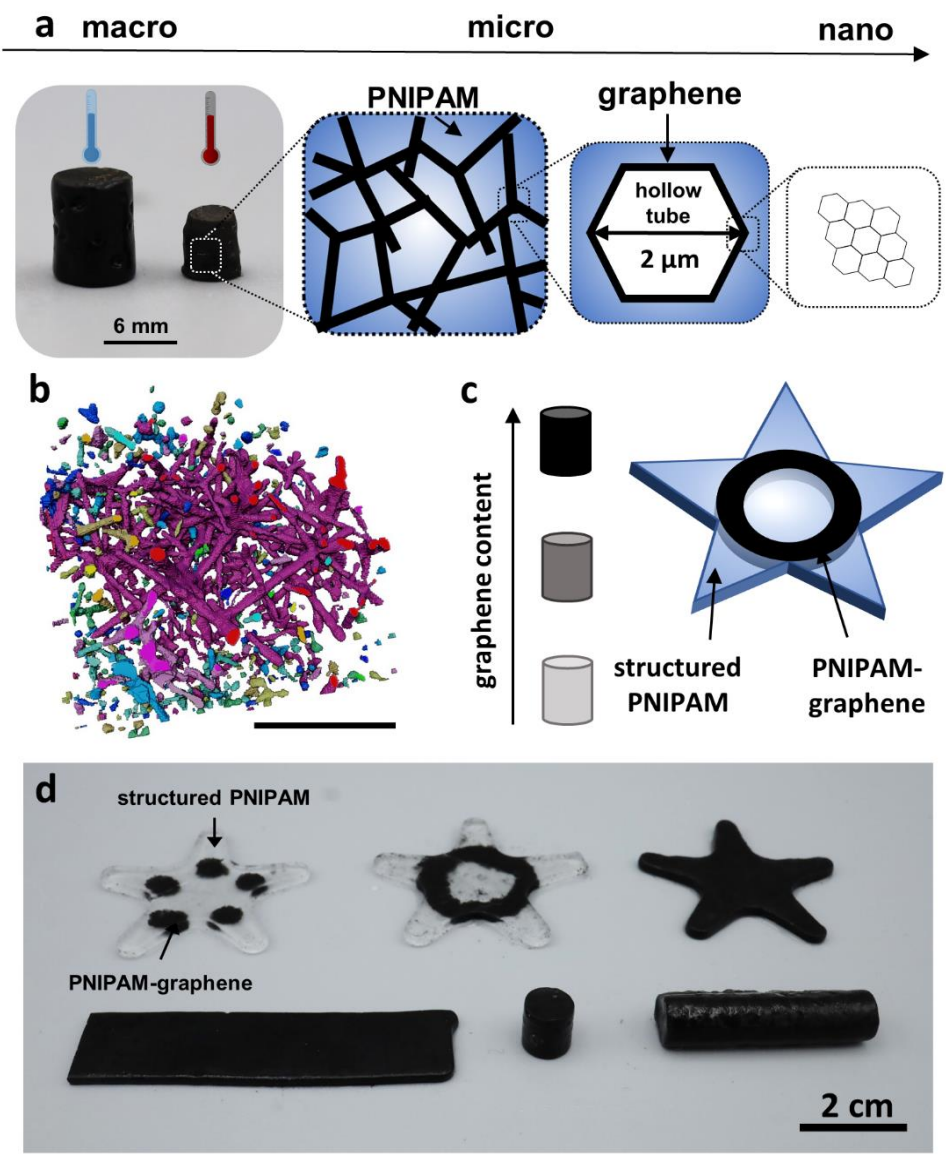

**Figure 1: Micro- and nanoengineered thermo-responsive poly(N-isopropylacrylamide)-exfoliated graphene (PNIPAM-EG) hydrogels**. (a) Combination of an interconnected hollow graphene microtube network and a PNIPAM hydrogel matrix. (b) 3D rendering of the microtube network obtained from micro computed tomography of PNIPAM-structured. Connected components are displayed in the same color. Scale bar is 200 µm. (c) The graphene content in PNIPAM-EG hydrogels is adjustable and can be applied to specific regions as a pattern. (d) PNIPAM-EG hydrogels with different shapes and geometries as well as with graphene patterns.

## 2.1 Mechanical properties and actuation performance: experimental and modelling

The mechanical properties as well as the deswelling and swelling kinetics of micro- and nanoengineered PNIPAM-EG hydrogels were investigated and compared to pure PNIPAM-structured and PNIPAM-bulk hydrogels. **Figure 2**a displays the compressive Young's modulus obtained within the initial strain of 5 % from the stress-strain curves (see **Figure S3**) for PNIPAM-bulk and for microstructured PNIPAM samples with different volumetric graphene concentrations (0 vol% to 0.35 vol%). PNIPAM-bulk has a compressive Young's modulus of 26 kPa. Introducing porosity to the hydrogel system does not load to a significant change in Young's modulus. The modulus of 25 kPa determined for PNIPAM-structured agrees with that predicted by a direct rule of mixture (24.6 kPa). Adding 0.35 vol% EG to the hydrogel matrix only increases the compressive Young's modulus to 36 kPa. Applying a direct rule of mixture here again, this suggests an equivalent Young's modulus of the graphene tube of 3.14 MPa (based on experimental data) and a linear dependence of the compressive Young's modulus with the graphene content, as observed in **Figure 2**a. The Young's modulus of the graphene wall is predicted to be around 3 GPa based on the ratio of the relative volumetric contents of pores and graphene in the hydrogel, and the graphene wall thickness is estimated to be around 32 nm (see **SI Note 1** for details). This reveals that the incorporation of a graphene network leads to a local stiffening of the microstructured PNIPAM hydrogels but maintains the global soft character, thanks to the hollow nature of the microtubes and low graphene content. The preservation of cyclic stability is demonstrated in **Figure 2**b, showing that the recovered height after 100 compression cycles at 20 % strain ranges between 94.5 % and 98 % for all sample types. The approach presented here enables the incorporation of an interconnected and continuous microtube network with an overall porosity of only 5.4 %, thereby maintaining the mechanical stability. Usually, adding porosity to a hydrogel matrix results in weakening of material stability[24]. Simultaneously, it has been reported previously, that such interconnected microtube networks can strongly enhance the water diffusion compared to bulk PNIPAM hydrogels, improving the deswelling kinetics.[20] **Figure 2**c and d show the deswelling (at 40 °C) and swelling (at 25 °C) curves as a function of time for PNIPAM-bulk, PNIPAM-structured and PNIPAM-EG (0.35 vol%), respectively. While PNIPAM-bulk only shows a deswelling of 16 % within 10 min, PNIPAM-structured deswells by 91 % at the same time, demonstrating that incorporation of an interconnected microtube network strongly enhances the water transport compared to bulk PNIPAM hydrogels, as previously reported.[20] The deswelling curves indicate that incorporating an additional graphene network in the microstructured hydrogels leads to a slightly reduced total shrinkage (84 %) within 10 min, compared to pure PNIPAM-structured. However, while the deswelling behavior is only slightly influenced by the addition of graphene, the swelling behavior (**Figure 2**d) is drastically improved. While the PNIPAM-EG hydrogels swell to their initial state within 3 h, the pure PNIPAM-structured hydrogels only reach 47 % of their initial mass in the same time (see **Figure S4** for photographs of the shrunken and swollen hydrogels). We suspect that this results from two effects:

First, the graphene layer alters the interface of the microtubes between the hydrogel and water-filled channel enhancing the transport through the modified skin layer and through the microtubes (see detailed explanation below). Second, due to its viscoelastic properties[38] the interconnected graphene network performs a restoring force when compressed. Indeed, the stress during swelling under the imposed axial confinement is expected to scale as the sum of two contributions: (1) the fluid pressure in the pores generated by capillary stress ($2 \gamma \cos(\vartheta) f r^{-1}$), where $\gamma$ is the surface tension of water, $\vartheta$ is the contact angle, $r$ is the channel radius and $f$ is the pore volumetric content) and (2) the elastic stress generated in the solid phases ($\eta E \varepsilon (1-f)$, where $E$ is the Young's modulus of the composite, $\varepsilon$ is the "potential" strain and $\eta$ is an efficiency less than one due to both softening at larger strain and viscoelasticity. The "potential" strain is the one that would be generated in the case of axial confinement removal, and can be estimated by the initial volume ($V_0$) and final volume (V): $\varepsilon = (V/V_0)^{1/3} - 1$. Accordingly, assuming $\eta = 1$, the elastic stress would be predicted to be 1.25, 5.60 and 7.68 kPa respectively for PNIPAM-bulk, -structured and -EG. However, since the capillary contribution for PNIPAM-bulk is expected to be negligible ($f = 0$), an efficiency $\eta = 0.37$ is estimated, resulting in an actuation stress of 0.48 kPa, as experimentally observed, see **Figure 2**e (and **Figure S5** for details on the measurement set-up). Assuming the same efficiency for PNIPAM-structured and -EG, the elastic stresses would be 2.07 kPa and 2.83 kPa, respectively. This suggests that graphene is enhancing the elastic recoil of the still soft hydrogel. Considering for graphene $2 \gamma \cos(\vartheta) \approx 60$ mJ m$^{-2}$[39] and assuming a microtube radius of 1 µm, the capillary stress is calculated to be 3.24 kPa for PNIPAM-EG, thus a total actuation stress (sum of elastic stress and capillary stress) of 6.07 kPa would be predicted for PNIPAM-EG. Experimentally the actuation stress is observed after 10 min to be of 18.70 kPa and asymptotically of 21.00 kPa (by best fitting with an exponential function (see **Figure S6**)), see **Figure 2**e. Taking this last value, the capillary stress is estimated to be around 18.17 kPa. This suggests the presence of equivalent nanoscopic channels activated by the graphene flakes on the microtube walls, having an equivalent characteristic radius of around 178 nm, when assuming a cylindrical geometry. The presence of these channels is further suggested by the actuation dynamics, as discussed in the next section and is strongly beneficial for improved actuation stress. To fit the observed actuation stress of 4.08 kPa for PNIPAM-structured (including the 2.07 kPa elastic stress), we assume $2 \gamma \cos(\vartheta) \approx 37$ mJ m$^{-2}$ [39], which is in agreement for the reported surface energy of PNIPAM (38.9 mJ m$^{-2}$ [40]). This suggests, that graphene is also enhancing the direct solid-water interaction.

Since reversible switching between two states is important for actuation applications, it is demonstrated here with cyclic swelling tests (**Figure 2**f), showing stable deswelling and swelling behavior over 10 cycles for PNIPAM-EG (0.35 vol%) as well as for PNIPAM-structured.

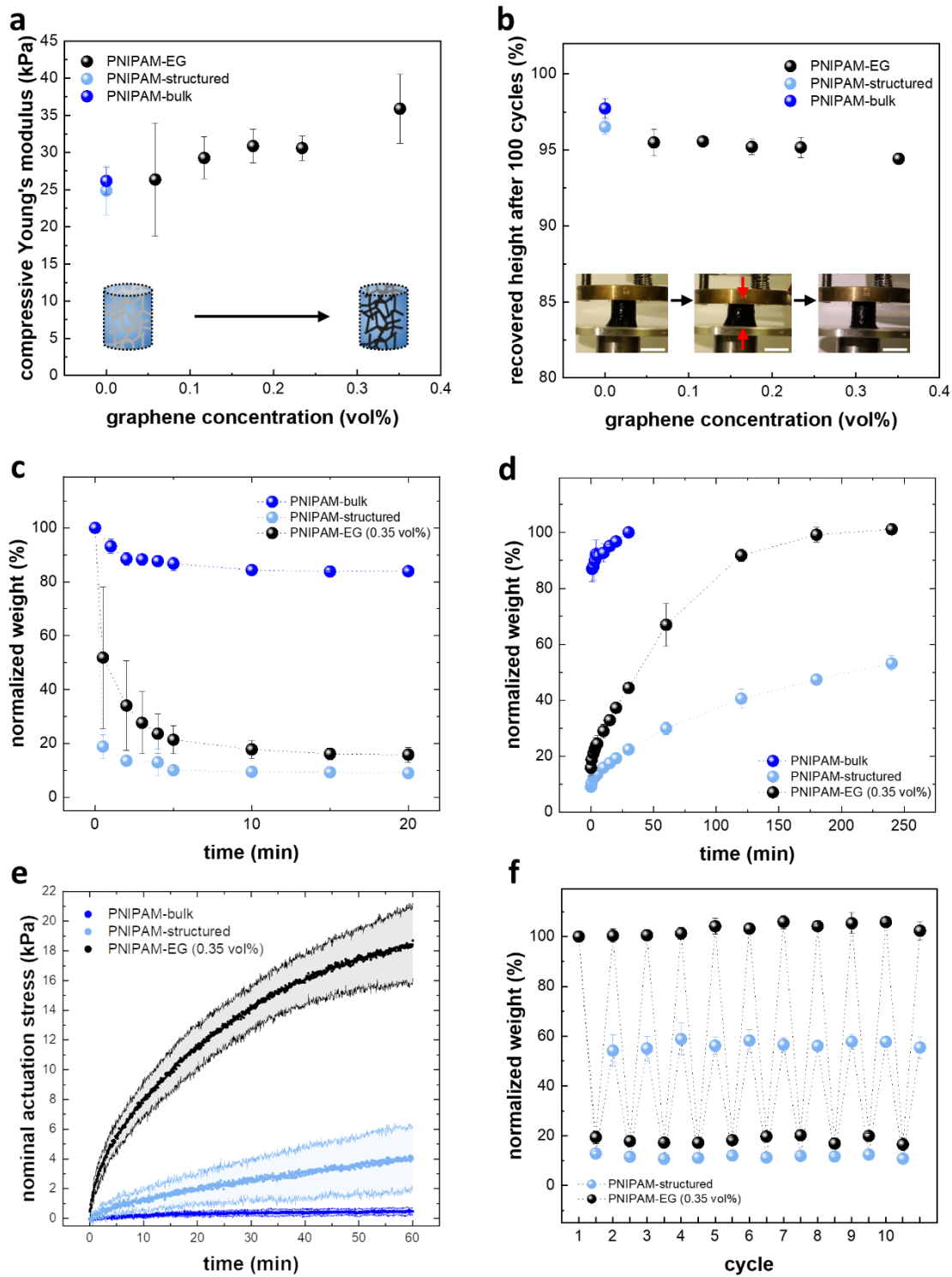

**Figure 2: Mechanical and actuation properties of micro- and nanoengineered PNIPAM hydrogels in comparison to bulk hydrogels.** (a) Compressive Young's modulus as a function of graphene concentration, obtained within the initial strain of 5 % from the stress-strain curves. (b) Cyclic stability in terms of the recovered height after 100 compression cycles at 20 % strain. Insets show images of cyclic compression test. Scale bar is 6 mm. (c) Deswelling curves and (d) swelling curves of PNIPAM-bulk, -structured and -EG (0.35 vol%). (e) Nominal actuation stress performed during swelling of PNIPAM-bulk, -structured and -EG (0.35 vol%). (f) Cyclic deswelling and swelling for 10 cycles of pure PNIPAM-structured and PNIPAM-EG (0.35 vol%). In all graphs error bars display the standard deviation (n=3).

## 2.4 Proposed model of the underlying mechanism of the actuation dynamics

The dynamics of hydrogel actuators are in general governed by water transport processes out of and into the polymeric network, limited by the poroelastic time scale of water diffusion. In the case of PNIPAM, the formation of a skin layer further restricts the actuation dynamics. For a more detailed understanding of the underlying mechanisms, that determine the deswelling and swelling behavior of the three investigated material systems (PNIPAM-bulk, PNIPAM-structured and PNIPAM-EG), the normalized weight is plotted against $t^{0.5}$ (**Figure 3**a), to visualize capillary and diffusion-based processes. For PNIPAM-bulk the deswelling within 20 min is limited to ~17 %, due to a dense skin layer that forms at the outer surface of the sample, drastically limiting further water diffusion out of the volume (**Figure 3**b). In contrast, the deswelling of PNIPAM-structured reaches ~80 % within 0.5 min. We suspect that the incorporation of an interconnected microtube network has a combined effect on the limitations due to poroelasticity and skin layer formation (**Figure 3**c). First, enhanced water transport out of the hydrogel is possible due to the geometrical assembly of the microtube network: (1) the distance between microtubes is small enough to ensure short diffusion pathways of water molecules from the hydrogel matrix to the microtubes, (2) the diameter of the microtubes (>2 µm) is large enough for unconfined water diffusion, (3) the microtubes are highly interconnected (~70 %). Second, we assume, that upon skin layer formation an increased amount of water is released as more interfacial hydrogel-water area, i.e. 188.7 cm² compared to 1.69 cm² (factor ~111, calculation details see **SI Note 2**) can be simultaneously dehydrated and transported through the microtubes, explaining the enhanced deswelling characteristics compared to the bulk. We suspect, that the microtubes close during the deswelling process. For PNIPAM-EG a slightly slower initial deswelling can be observed, which can be attributed to a counter force exerted by the interconnected graphene network, still reaching 49 % deswelling within 0.5 min. Due to improved solid-water interactions inside the graphene-coated microtubes, we assume that for PNIPAM-EG the microtubes remain partially open and water-filled (**Figure 3**d), limiting the total deswelling to 84 %. With respect to the swelling behavior, PNIPAM-bulk and PNIPAM-structured show an equal slope (**Figure 3**a). This indicates that the process is not limited by simple diffusion through the hydrogel matrix, but rather to the transport through the dense skin layer, as otherwise the geometrical factor should lead to different slopes in the $t^{0.5}$-plot, due to longer pathways. Thus, we assume, that the water transport through the interfacial skin layer is the rate-limiting process, strongly impairing the swelling of both hydrogel systems. A completely different behavior is observed for the swelling of PNIPAM-EG hydrogels. Initially the curve also shows a linear behavior with a similar slope as PNIPAM-bulk and -structured, but deviates strongly in the following part. As mentioned above, we suspect, that the graphene-coated microtubes still contain water in the deswollen state. This reduces the water diffusion pathways, resulting in a reduction of the poroelastic diffusion time scale and homogeneous swelling of the PNIPAM-EG hydrogels. Additionally, the incorporation of a graphene network into the hydrogel matrix creates a hydrogel-graphene-water interface inside the microtubes (**Figure 3**d). We suspect that the altering of the interface influences the water diffusion through the skin layer, resulting in faster swelling kinetics. A possible effect could be that graphene sheets perforate the interfacial skin layer, thereby creating the nanoscopic channels previously discussed and pathways for facilitated water transport through the interface. At the same time, the graphene network performs an increased actuation stress, as shown in **Figure 2**e, which could also lead to facilitated swelling, as previously quantified.

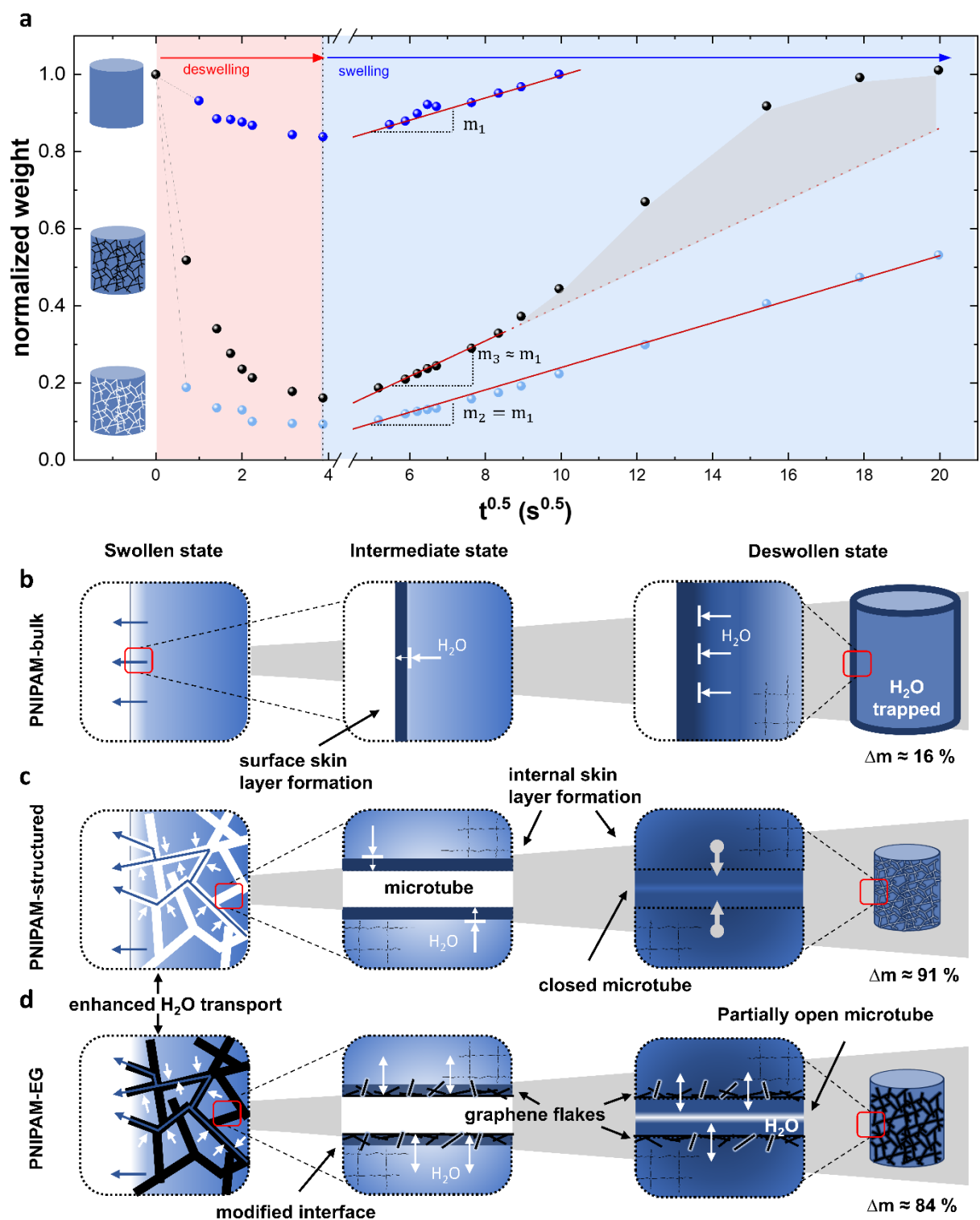

**Figure 3: Proposed model for the volume change upon phase transition across the LCST of micro- and nanoengineered PNIPAM hydrogels.** (a) Deswelling and swelling curves as a function of $t^{0.5}$ of PNIPAM-bulk, PNIPAM-structured and PNIPAM-EG (0.35 vol%). (b) The slow and limited deswelling of PNIPAM-bulk hydrogels is due to the poroelastic time limit and skin layer formation around the sample which restricts water transport out of the hydrogel. (c) Incorporating a network of interconnected microtubes into the hydrogel matrix enhances water transport out of the hydrogel volume, thereby improving the poroelastic limit. At the same time the hydrogel-water interface area is increased by a factor of ~111, enhancing water transport through the skin layer. Both effects result in improved deswelling of PNIPAM-structured hydrogels, while the swelling is still limited, assuming that the microtubes close during deswelling. (d) Further coating of the microtubes by graphene

additionally modifies the hydrogel-water interface. Thus, the microtubes might remain open and partially water-filled during deswelling which improves the subsequent swelling of PNIPAM-EG hydrogels. Additionally, the graphene flakes might perforate the skin layer, facilitating water transport into the hydrogel volume through nanoscopic channels (as also suggested by modelling), resulting in enhanced swelling.

## 2.4 Photothermal heating

Next to improvement of performance characteristics, precise control over trigger options is required for hydrogel-based actuators. While the standard method to induce actuation of thermo-responsive hydrogels is to change the temperature of the surrounding environment, we here show untethered photothermal as well as rapid Joule heating by utilizing the unique properties of graphene, i.e. broadband light absorption and electrical conductivity. Light-induced heating was studied for PNIPAM hydrogels with different graphene concentrations (0.0004 vol% EG – 0.18 vol% EG) (**Figure 4**a). Illumination with white light results in a local light-to-heat conversion of the PNIPAM-EG hydrogel, due to broadband light absorption of graphene, as demonstrated schematically and by thermography in **Figure 4**b. The maximum surface temperature of the hydrogels is shown in **Figure 4**c as a function of time for the different graphene concentrations. While the pure PNIPAM-structured only heats up by approx. 1 °C within 20 s of illumination with 0.44 W cm$^{-2}$, all PNIPAM-EG hydrogels with 0.0021 vol% EG and higher concentrations heat up above the lower critical solution temperature of 32 °C within a few seconds. The initial heating rate (**Figure 4**d) increases with higher volumetric loadings of EG and reaches values up to 11.6 °C s$^{-1}$ for PNIPAM-EG (0.18 vol%). **Figure 4**e shows the normalized weight of hydrogels with different graphene concentrations after illumination with 0.44 W cm$^{-2}$ for 20 s, highlighting a weight reduction of ~30 % - 40 % for PNIPAM-EG with 0.0042 vol% and higher concentrations (see **Figure S7** for images of light-induced deswelling). The influence of the illumination time on the deswelling of the hydrogels was measured for PNIPAM-EG (0.06 vol% EG), see inset of **Figure 4**e, showing that shorter illumination times lead to less deswelling. For low concentrations of EG within the hydrogel matrix, a large fraction of the incoming light is transmitted through the structure, rather than absorbed by the graphene, resulting in a low heating rate of the hydrogel. With increasing EG concentration, more light is absorbed leading to higher temperatures. For the highest measured EG concentration (0.18 vol%), light penetration is limited as light is directly absorbed in the surface region leading to a high surface temperature. However, due to the high water content, heat transport through the hydrogel matrix is limited, resulting in an inhomogeneous deswelling of the sample, see **Figure 4**e. The increase in temperature upon illumination and thus, the deswelling, can also be controlled by adjusting the light intensity, as demonstrated for PNIPAM-EG (0.06 vol%) in **Figure 4**f. A light intensity of 0.44 W cm$^{-2}$ results in fast heating with an initial heating rate of 9 °C s$^{-1}$, while a lower intensity of 0.162 Wcm$^{-2}$ leads to an initial heating rate of 1.5 °C s$^{-1}$. The findings demonstrate that the temperature increase of PNIPAM-EG can be controlled by varying graphene concentration, light intensity and illumination time, which in turn facilitates control over the deswelling kinetics. Further, photothermal heating is a beneficial tool for any controlled actuation requiring local heating and phase change of the hydrogel, as it is restricted to the area of illumination as the water-containing hydrogel limits the heat transfer throughout the material volume. **Figure 4**g demonstrates the local heating of a star-shaped PNIPAM-EG hydrogel resulting in the defined movement of only one of the star's arms within 4 s of illumination (see **Video S1**).

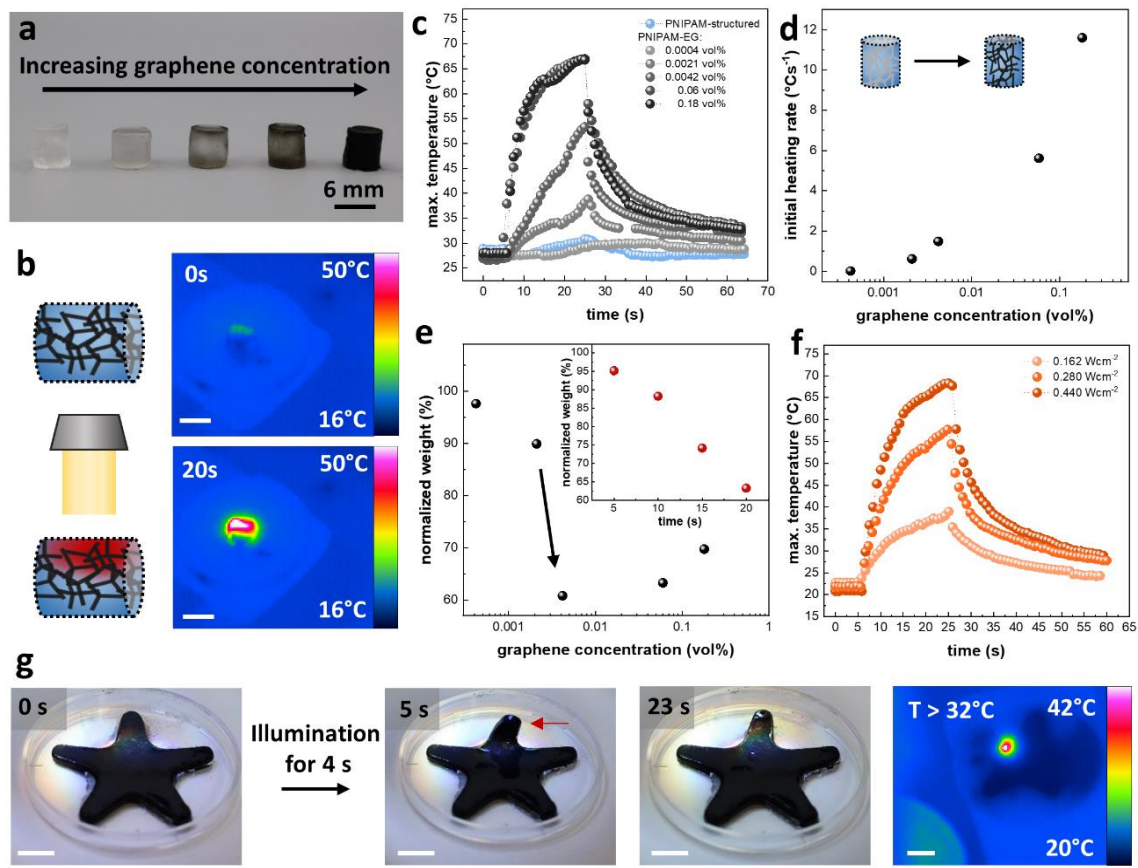

**Figure 4: Photothermal heating of micro-and nanoengineered PNIPAM-EG hydrogels with varying graphene content.** (a) Photograph of PNIPAM hydrogels with increasing graphene concentration, from left to right: 0 vol%, 0.0004 vol%, 0.0021 vol%, 0.0042 vol%, 0.06 vol% EG. (b) Schematic representation and thermograms of the photothermal heating of PNIPAM-EG. Scale bar is 6 mm. (c) Temperature curves of PNIPAM hydrogels with varying EG content, upon illumination with white light (0.44 W cm$^{-2}$) and (d) corresponding initial heating rate as a function of graphene concentration. (e) Normalized weight as a function of graphene concentration after illumination with white light (0.44 W cm$^{-2}$) for 20 s. The inset shows the normalized weight as a function of illumination time for PNIPAM-EG (0.06 vol%). (f) Temperature curves of PNIPAM-EG (0.06 vol%) for different light intensities. (g) Photographs and corresponding thermogram of a star-shaped PNIPAM-EG hydrogel, locally illuminated for 4 s with white light (0.44 W cm$^{-2}$). Scale bar is 5 mm.

**2.5 Joule heating**

In contrast to locally defined heating by photothermal conversion, Joule heating (also referred to as resistive heating) can be used to homogeneously and rapidly heat the entire hydrogel volume in a controlled and defined manner. The interconnected conductive EG network forms pathways of 'microwires' penetrating the entire hydrogel matrix, while the global elasticity of the hydrogel matrix is maintained (see **Figure 3**a). Previously, it has been demonstrated, that graphene-based aerogels and framework structures provide an excellent platform for the development of innovative energy transducers, e.g. electricity into heat, due to their high conductivity, low density and low volumetric heat capacity.[41] Here, we utilize those properties to rapidly heat our thermo-responsive hydrogel systems to enable a controlled and defined actuation. Most importantly, the establishment of a good electrical contact to the PNIPAM-EG hydrogels is crucial for a homogeneous and efficient heating. To ensure that most of the potential drops across the network and is converted into heat, the contact resistance must be lower than the resistance of the graphene network. For this, copper foils and silver wires were attached to both sides of the cylindrical graphene coated ZnO templates using conductive silver paste before filling the template with hydrogel precursor solution during synthesis. Thereby, a specific conductivity of up to 2.16 S m$^{-1}$ could be obtained for PNIPAM-EG (0.29 vol%). Additionally, the cross-linker amount was increased to 16.1 % to reduce the swelling of the hydrogels, and thus, prevent the contact to rupture from the sample during deswelling and swelling. Characterizations of the hydrogels with higher cross-linking can be found in the supplementary information (**Figures S8 and S9**), showing the same trends as PNIPAM-EG hydrogels with lower cross-linking. Thermography (**Figure 5**a) shows that applying a voltage results in a heating of the entire volume of the PNIPAM-EG hydrogels. In turn, the homogeneous heating of the material leads to a uniform deswelling of the entire hydrogel volume, as depicted in **Figure 5**b for a PNIPAM-EG (0.29 vol%) hydrogel (see **Video S2**). By varying the applied voltage, the heating of the material can be controlled, as shown in **Figure 5**c for a PNIPAM-EG (0.35 vol%) hydrogel, demonstrating that higher applied voltages result in higher heating rates (0.6 °C s$^{-1}$, 1.7 °C s$^{-1}$, 2.7 °C s$^{-1}$ for 15 V, 20 V, 25 V, respectively). The resulting length change of the hydrogels also increases with higher applied voltages (**Figure 5**d), reaching a maximum length change of 22 % for a voltage of 25 V. Additionally, **Figure S10** displays the course of the current of the PNIPAM-EG (0.35 vol%) hydrogel during Joule heating. Homogeneous heating of the hydrogel results in deswelling, and thus, in a compression of the graphene network, bringing the graphene sheets closer together, leading to an increase in current. Further, Joule heating can be performed in cyclic manner (see **Video S2**). The normalized length change for several heating cycles is shown in **Figure 5**e with an applied voltage of 25 V for 1.5 min and subsequent swelling for 6 min. The length changes about 17 % and almost reaches the initial state after swelling. Again, the current (**Figure 5**f) increases during electrical heating associated with the compression of the graphene network upon deswelling of the hydrogel.

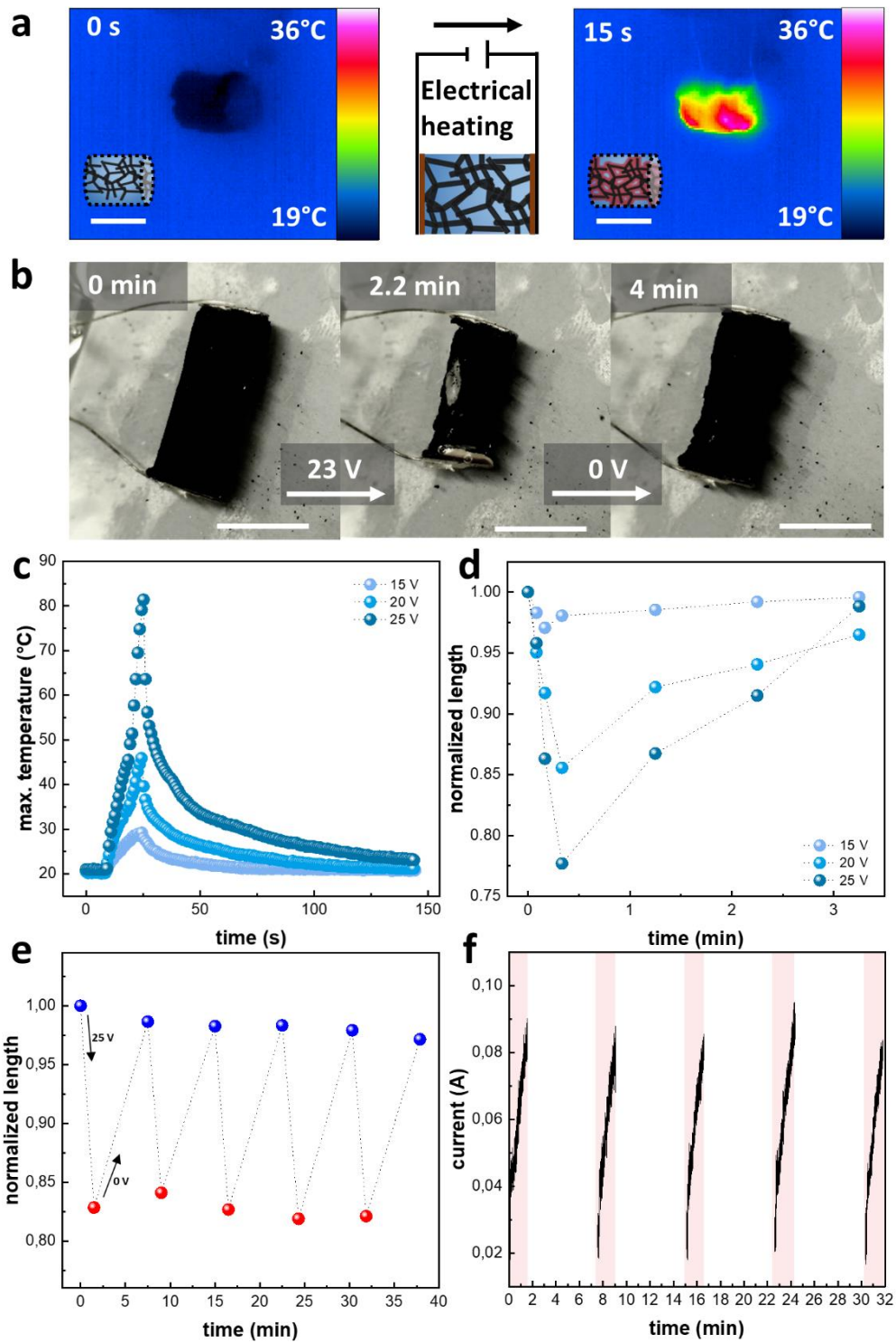

**Figure 5: Volumetric Joule heating of micro-and nanoengineered PNIPAM-EG hydrogels.** (a) Thermograms of electrically heated PNIPAM-EG hydrogel. Scale bar is 6 mm. (b) Photographs of PNIPAM-EG (0.29 vol%) hydrogel heated with 23 V in water. Scale bar is 6 mm. (c) Temperature curves and (d) corresponding normalized length as a function of time for different applied voltages for a cylindrical PNIPAM-EG (0.35 vol%) sample with 6 mm height. (e) Cyclic deswelling and swelling of an electrically heated PNIPAM-EG (0.35 vol%) hydrogel (10 mm height) in water, in terms of the normalized length as a function of time and (f) corresponding current curves.

## 3 Actuator designs

As demonstrated, the precise micro- and nanostructuring of the PNIPAM-EG hydrogels leads to improved actuation dynamics and performance, while also enabling internal heating of the hydrogels. To demonstrate the suitability and variability of the microengineered hydrogels for soft actuators, different actuator designs and concepts were developed and studied. For an untethered, light-controlled actuation a beam actuator and a bilayer-structured gripper were designed. The beam actuator (**Figure 6**a) can be actuated by local heating with light, leading to a local temperature increase above the LCST within 20 s (**Figure 6**b). Local deswelling then results in a lift of the attached weight, as demonstrated in **Figure 6**c for two different weights. The motion can be performed in cyclic manner in water (T<LCST), as shown in **Video S3** and **Figure 6**d, displaying the height of the lifted weight (1 g)**.** Further, the presented microstructuring approach facilitates the fabrication of a bilayer-structured gripper for untethered, light-controlled actuation. The gripper consists of a microstructured PNIPAM-EG layer and an unstructured PNIPAM layer on top (**Figure 6**e). The two layers consist of the same hydrogel (PNIPAM). Thus, the bilayer structure can be prepared by filling the t-ZnO-EG template, used for microstructuring, with the hydrogel precursor solution and pouring excess solution onto the template to create the additional unstructured PNIPAM layer. In contrast to other approaches [42–45], here only one hydrogel system and polymerization step is needed without any further chemical modification. The introduction of strain in the bilayer structure results from two effects: firstly, the PNIPAM-EG layer heats up upon illumination, unlike the pure PNIPAM layer, and secondly, the microstructured PNIPAM-EG layer additionally shows a stronger deswelling than the unstructured PNIPAM layer. The gripper can be actuated by global illumination, leading to heating and strong deswelling of the microstructured PNIPAM-EG layer, whereas the unstructured PNIPAM layer deswells only slightly, resulting in a strain and final grasping of an object within 80 s, shown in **Figure 6**f and **Video S**3. The object can be released instantly by placing the gripper in water (T<LCST) (**Figure 6**g).

To demonstrate the application of volumetric Joule heating an electrically powered gripper was designed (**Figure 6**h). The actuation is shown in the image series in **Figure 6**i and **Video S3**. Joule heating with 25 V results in deswelling of the PNIPAM-EG hydrogel and the gripper closes via force transmission and thus, an object can be moved. Upon cooling and swelling of the hydrogel in water, the gripper opens again and the object is released.

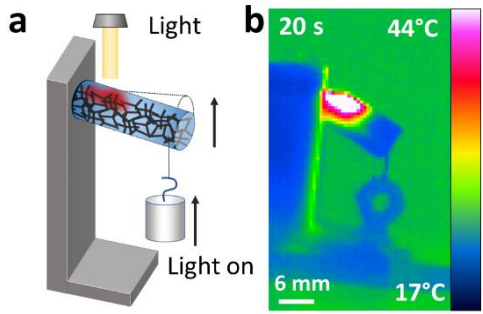
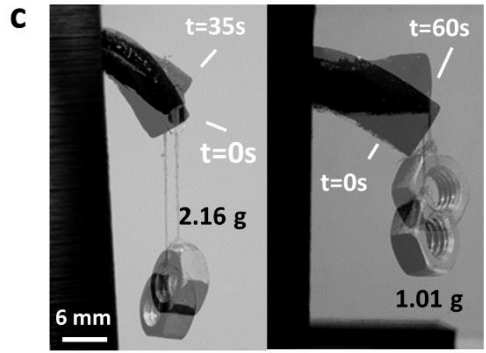
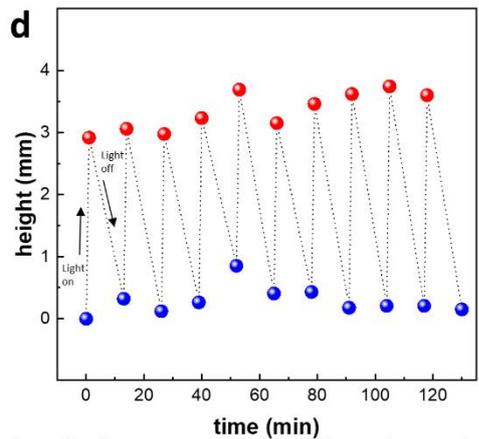
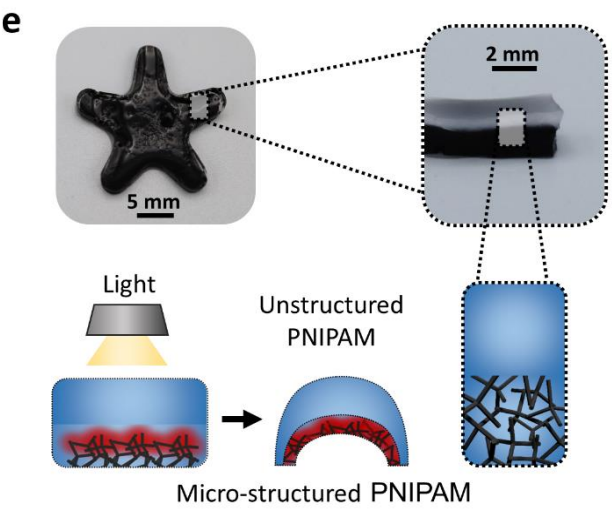
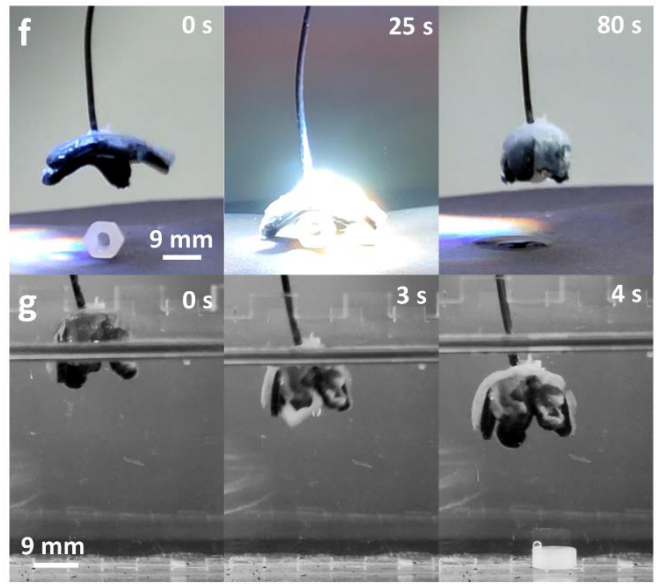
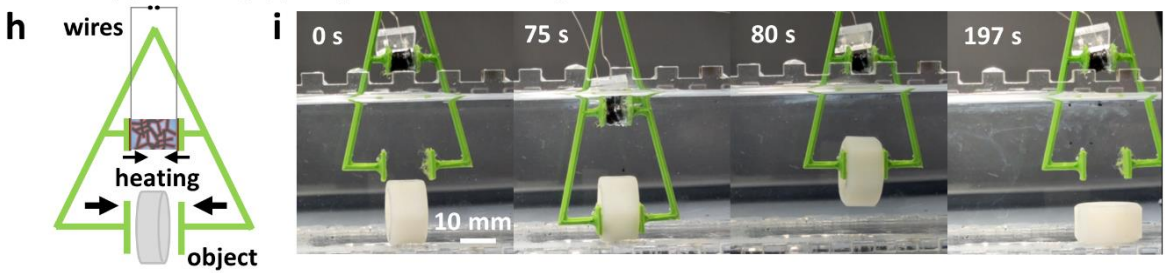

**Figure 6: Hydrogel actuator designs and concepts based on micro-and nanoengineered PNIPAM-EG hydrogels.** (a) Schematic of concept for an untethered light-controlled beam actuator. (b) Thermogram of beam actuator, illuminated with white light (0.44 W cm$^{-2}$). (c) Merged photographs of beam actuator, lifting attached weights. (d) Height of the attached weight as a function of time for multiple actuation cycles of the beam actuator in water. (e) Photographs and schematic of an untethered light-controlled gripper, consisting of a microstructured PNIPAM-EG layer with an unstructured PNIPAM layer on top. (f) Photo series of grasping and (g) releasing an object by the light-controlled gripper. (h) Schematic of an electrically powered gripper by volumetric Joule heating. (i) Photo series of grasping and releasing an object by the electrically powered gripper.

**4 Conclusion**

In summary, we presented a micro- and nanoengineering approach for the preparation of thermo-responsive hydrogel-based soft actuators with improved actuation stress and dynamics and overall performance. Incorporation of an interconnected microtube graphene network into a PNIPAM matrix overcomes the limitations related to poroelasticity and skin layer formation in bulk PNIPAM hydrogels. This leads to an improvement in actuation dynamics of up to ~400 %, while the introduction of only 5.4% porosity maintains mechanical stability. The combination of interconnected graphene network and hydrogel matrix results in increased actuation stress (+ ~4000 %), without compromising the global softness of the PNIPAM matrix. Tailoring the graphene concentration within the material system provides easy control over response times and enables the application in soft actuation systems. Besides the standard actuation method for thermo-responsive hydrogels, the presented PNIPAM-EG hydrogel system allows untethered, light-triggered as well as electrically-powered and controlled actuation for soft actuators. The concept is transferable to other hydrogel systems to overcome poroelastic limitations, while allowing the incorporation of microtube networks composed of other nanomaterials, thereby extending their functionalities for application as soft sensors and actuators.

**Materials and Methods**

**Microstructured poly(N-isopropylacrylamide) (PNIPAM) hydrogels** were prepared by using sacrificial templates from tetrapodal zinc oxide (t-ZnO). T-ZnO powder was prepared by a flame transport synthesis, described in detail elsewhere[46,47]. The loose t-ZnO microparticles were pressed into templates using metal molds with desired geometry (cylindrical (d=6 mm, h=6/10/13/20 mm) or star-shaped (h=2 mm)). In this work sacrificial templates with 0.3 g cm$^{-3}$ t-ZnO were prepared and then sintered at 1150 °C for 5 h to obtain stable, interconnected networks. The desired density was achieved by using the appropriate amount of t-ZnO for a given volume of the mold. For the preparation of microstructured conductive poly(N-isopropylacrylamide)-exfoliated graphene (PNIPAM-EG) hydrogels, the ZnO tetrapods were coated with a thin layer of graphene by a wet-chemical infiltration method, reported in a previous publication[38]. For this, the entire free volume of the t-ZnO templates was filled with an aqueous dispersion of electrochemically exfoliated graphene (EG) (1.4 mg mL$^{-1}$) and dried on a heating plate at 40 °C for 4 h, resulting in a homogenous coating of the ZnO tetrapod arms. PNIPAM-EG hydrogels with different graphene concentrations, ranging from 0.0004 vol% to 0.35 vol%, were prepared by either diluting the EG dispersion or performing multiple infiltration steps. Samples intended for Joule-heating were then contacted using silver paste (Acheson 1415) and copper foil. Subsequently, a precursor solution of the hydrogel (PNIPAM) was prepared by mixing ammonium persulfate (APS) (1:10 solution, 100 µl), distilled water (dH2O) (1000 µl or 750 µl), 2%-N,N'- methylenebisacrylamide (BIS) (500 µl or 750 µl), 20%-N-isopropylacrylamide (NIPAM) (3000 µl) and N,N,N',N'-tetramethylethylenediamine (TEMED) (46 µl) (all chemicals purchased from Sigma Aldrich) in a beaker and infiltrated into the templates to fill up the entire free volume. Bulk hydrogels were prepared by pouring the solution into cylindrical molds with diameter and height of 6 mm. Bilayer-structured samples were prepared by placing a star-shaped ZnO template in a well plate and filling the well with an additional layer (ca. 2 mm) of hydrogel precursor solution. All samples were prepared with 10.8 % cross-linker (BIS) except samples for Joule heating which were fabricated with 16.1 % cross-linker. After 5 h of polymerization the samples were either placed in distilled water (bulk samples) or in diluted hydrochloric acid (HCL, 0.5 M) for 3 days to etch the t-ZnO resulting in microstructured samples with an overall porosity of 5.4 %. The porosity was calculated from the volume occupied by t-ZnO $V_{\text{t-ZnO}}$ and the volume of the template $V_{\text{template}}$ according to equation (1):

$$Porosity\ (\%) = \frac{V_{\text{t-ZnO}}}{V_{\text{template}}} \tag{1}$$

Afterwards, all samples were thoroughly washed with distilled water by exchanging the water twice a day for at least 5 days to remove access monomers and acid.

**For micro computed tomography (microCT) analysis** a PNIPAM sample was cut to a size of approx. 1.6 x 1.6 x 3.75 mm³ and secured by mounting on a 0.2 mm diameter drill. The sample was then placed in a Kapton tube of 1.6 mm diameter with water to ensure its hydration. Both ends of the Kapton tube were sealed using radiotransparent resin hardened using blue light (easyform LC, DETAX GmbH & Co. KG, Ettlingen, Germany). The sample was at the P05 microtomography beamline IBL operated by Helmholtz-Zentrum Hereon at the PETRA III storage ring at Deutsches Elektronen-Synchrotron (DESY) in Hamburg. The photon energy was set to 30 keV. A 5120 x 3840 pixel KIT CMOS camera was used at 10X magnification, resulting in an effective pixel size of 0.64 µm. To achieve sufficient propagation-based phase contrast, the sample-to-detector distance was set to 500 mm. The sample movement was minimized further by performing a flyscan with 2000 projections (with 20 dark and 300 flat field images) and an exposure time of 200 ms, thus leading to an overall scan time of approx. 4 minutes. The tomogram reconstructed with bin 2 using a customized reconstruction tool[48] that uses the Astra Toolbox[49,50] in Matlab R2021b (The MathWorks Inc., USA).

Due to high image noise from fast imaging and ring artefacts, the image was filtered using a custom filter to remove ring artefacts implemented in Python (Anaconda 3, version 5.2), followed by iterative non-local means filter[51] with an isotropic search radius of 6 voxel and 4 iterations. As some sample regions displayed minor movement and remaining artefacts, a region-of-interest (ROI) of 0.355 x 0.355 x 0.357 mm³ was selected for analysis. The trainable WEKA plugin[52] in Fiji/ ImageJ[53] was used for the segmentation of the pores. The connectivity of the sample was determined in Avizo 2021.1 (FEI SAS,

Thermo Scientific, France) allowing for a minimum component size of 27 voxels and corner connectivity. Finally, the connectivity was calculated as the fraction of the volume of the largest connected pore divided by the overall pore volume.

**Deswelling and swelling curves** were recorded by storing the cylindrical samples (d=6 mm, h= 6mm) in a water bath at 40 °C or 25 °C, respectively, and weighing the samples at distinct time points. For cyclic tests, the samples were weighed after 15 min in a water bath at 40 °C. Subsequently, the samples were kept in a water bath at 25 °C for 3 h and then weighed. Triplicates were measured.

**The cyclic compression tests** of swollen cylindrical samples (d=6 mm, h= 6mm) were performed using a customized set-up (Märzhäuser Wetzlar HS 6.3 micromanipulator and force sensor (burster, type 9235/36)), controlled by a LabView program. Compression tests were performed in air at a strain rate of 2 %s$^{-1}$. Triplicates were measured.

**The measurement of the nominal actuation stress during swelling** was conducted with the same set-up as for the cyclic compression tests. First, the cylindrical samples (d=6 mm, h= 6mm) were dehydrated in a water bath at 40 °C and then placed between the upper and lower plate in a water bath at 25 °C. The plates stayed at a constant position while the exerted force was measured.

**The light-induced heating** was characterized with a self-built set-up consisting of a beamer (Acer, DLP Projector, DNX0906) and a Fresnel lens (from an overhead projector) to focus the light onto the sample. The light intensity was controlled using a power point presentation with different grey scales and measured by a photometer (Thorlabs Photometer, S425C-L with Interface PM100USB and Thorlabs Opitcal Power Monitor). For determination of light-induced deswelling, the samples were weighed before and after illumination. Characterization of light-induced heating was performed in air with cylindrical samples (d=6 mm, h= 6 mm).

**For Joule heating** silver wires were attached to the copper foil at both sides of the PNIPAM-EG hydrogels using conductive silver paste (Acheson 1415). A constant voltage was applied using a USB-2537 High-Speed DAQ Board (Measurement Computing Corporation, USA) and a self-written LabView program with an EA-PS 2042-10B (EA Elektro-Automatik, Germany) as power supply. Simultaneously, voltage and current were measured as a function of time. Characterization of the temperature profiles and respective deswelling were performed in air with cylindrical samples (d=6 mm, h= 6 mm), while cyclic tests were performed in distilled water with cylindrical samples with a height of 10 mm or 13 mm. The specific conductivity $\sigma$ was calculated from the current $I$, the voltage $U$, sample length $l$ and sample cross-section $A$ according to the following formula:

$$\sigma = \frac{I \cdot l}{U \cdot A} \qquad (1)$$

**Temperature recording** was done with an IR-thermo camera (IRBIS) in combination with IRBIS professional software.

**Actuator designs:**

For the beam actuator a cylindrical PNIPAM-EG (0.18 vol% EG, 10.8 vol% cross-linker) hydrogel with 2 cm height was fixed at one end while a weight was attached to the other end. Actuation was performed by light-induced heating with the set-up described above. Cyclic actuation was investigated in distilled water, whereas lifting of ~2 g was performed in air.

The light-controlled gripper was fabricated using a bilayer-structured star-shaped sample (0.18 vol% EG, 10.8 vol% cross-linker) attached to a thin rod. To achieve a grasping movement, the actuator was illuminated with 0.44 W cm$^{-2}$ using the beamer set-up in air. Releasing of the object was done by immersing the gripper in distilled water (T<LCST).

The electrical gripper was 3D-printed using polylactic acid. Silver wires were attached to a PNIPAM-EG hydrogel (0.35 vol% EG, 16.1 % cross-linker) at the copper foil with conductive silver paste (Acheson 1415). The hydrogel was then fixed at the desired position with superglue. For actuation a voltage of 25 V was applied. The experiment was performed in water.

**Declaration of Competing Interest**

The authors declare that they have no known competing financial interests or personal relationships that could have appeared to influence the work reported in this paper.

**Acknowledgments**

The authors thank Dr. Jürgen Carstensen for fruitful discussions. MH, RA and FS gratefully acknowledge funding by the German Research Foundation (DFG) through the Research Training Group "Materials for Brain" (Project P3) and grant SCHU 3506/4-1. LMS, ASN, XF, RA, FS, and the Humboldtian NMP gratefully acknowledge funding by the EU through the GrapheneCore3 881603 Project. BZP and FS acknowledge support by the Kiel Nano Surface and Interface Science (KINSIS).

**Data availability**

The data that support the findings of this study are available from the corresponding authors upon request.

# Supporting Information

## Fabrication of PNIPAM-EG hydrogels

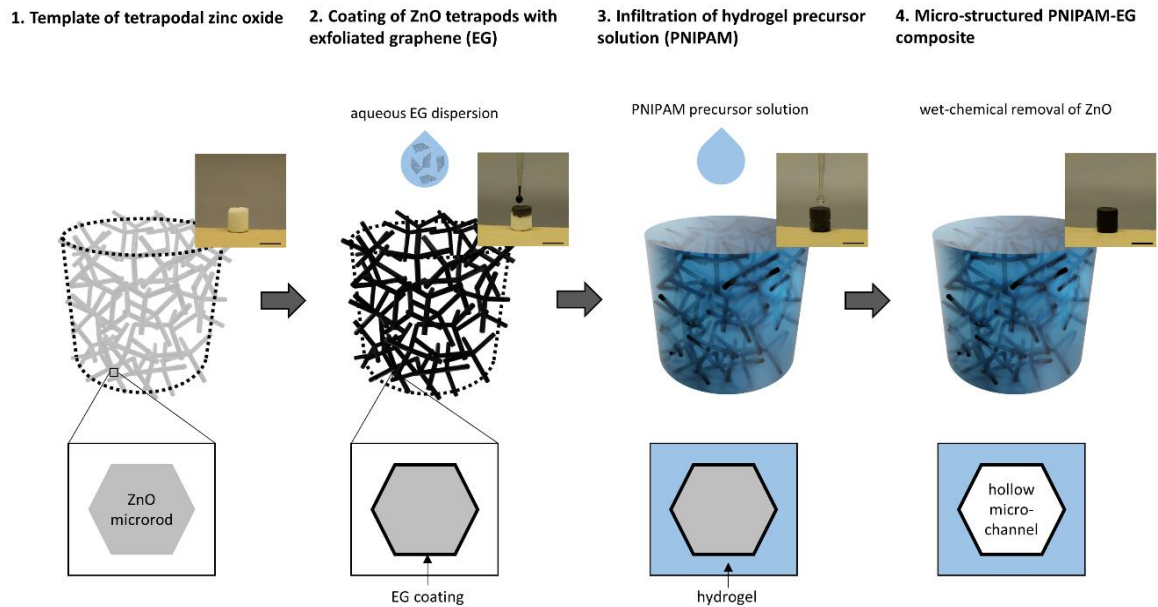

**Figure S1: Fabrication of PNIPAM-EG hydrogels.** 1. Sacrificial template of tetrapodal zinc oxide (t-ZnO). 2. Wet-chemical coating of the ZnO tetrapods with exfoliated graphene (EG) by drop-casting aqueous dispersion onto the template, followed by drying on a heating plate. 3. Filling of the entire free volume of the ZnO template with poly(N-isopropylacrylamide) (PNIPAM) precursor solution. 4. After polymerization, the hydrogels are stored in hydrochloric acid to remove the ZnO, resulting in an interconnected network of hollow micro-channels.

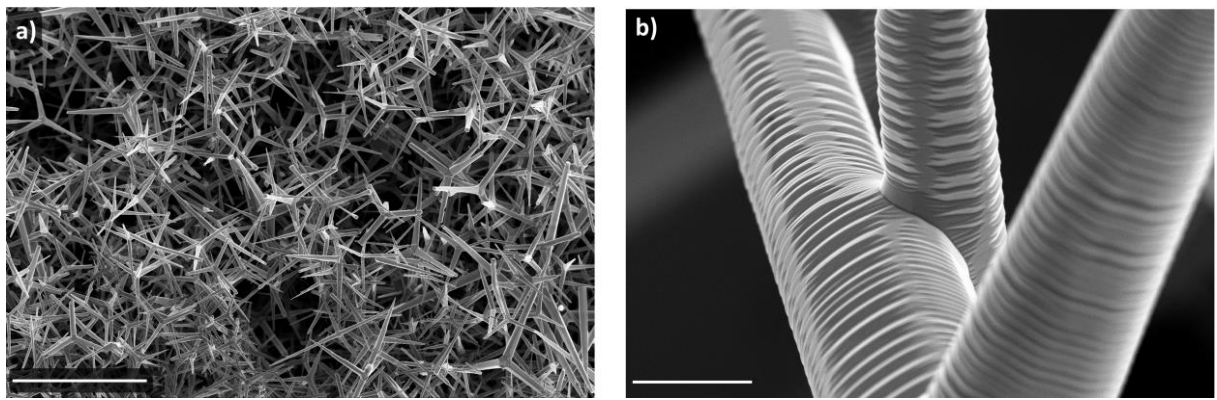

**Figure S2: Scanning electron microscopy images of tetrapodal zinc oxide (ZnO).** a) Network structure of interconnected ZnO tetrapods (scale bar is 100 μm). b) Interconnection between two ZnO tetrapods (scale bar is 2 μm).

**Mechanical properties of PNIPAM-structured and PNIPAM-EG**

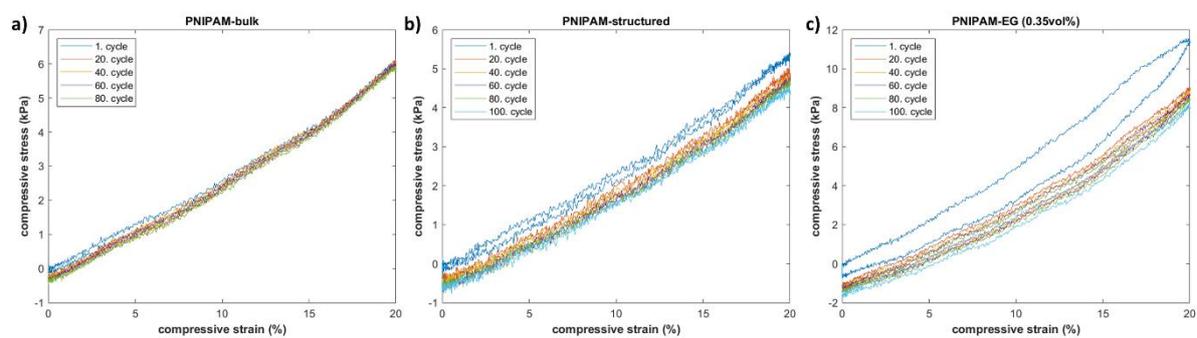

**Figure S3:** Compressive stress-strain curves of (a) PNIPAM-bulk, (b) PNIPAM-structured and (c) PNIPAM-EG (0.35 vol%) for 100 cycles.

**Deswelling and swelling properties of PNIPAM-structured and PNIPAM-EG**

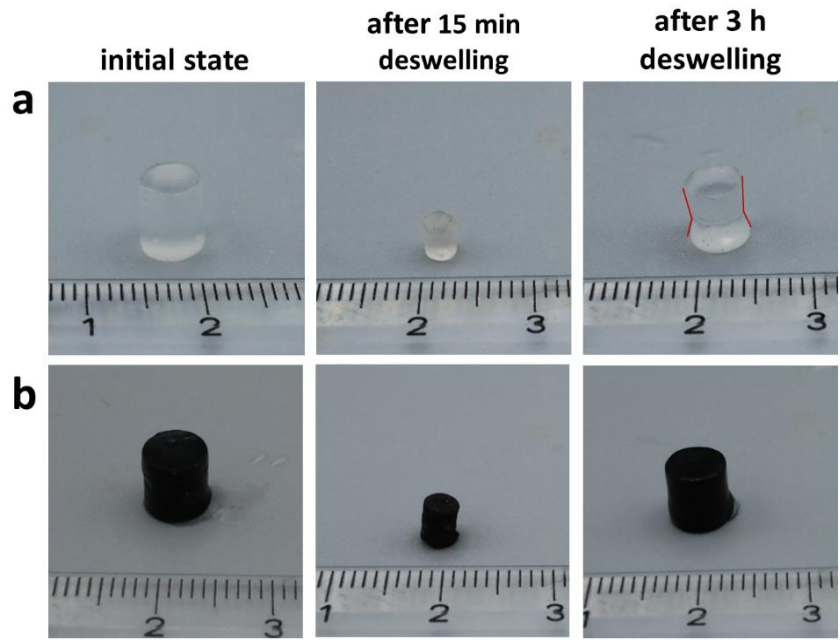

**Figure S4:** Photographs of (a) PNIPAM-structured and (b) PNIPAM-EG (0.35 vol%) in the initial state, after 15 min of deswelling in a water bath at 40 °C and after 3 h of swelling in a water bath at 25 °C.

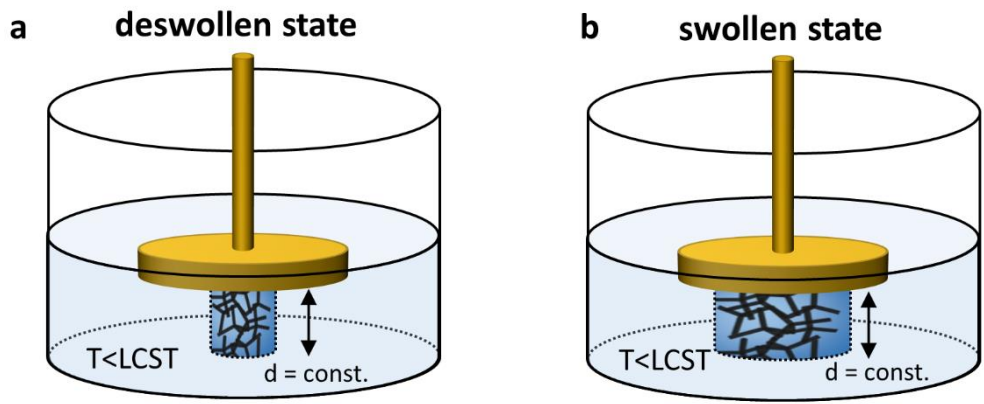

**Figure S5: Schematic of the set-up for measurement of the swelling force.** (a) The sample deswollen sample is immersed in a water bath with T<LCST and fixed by a plate. (b) During swelling of the sample, the distance of the plate is kep constant and the exerted force is measured.

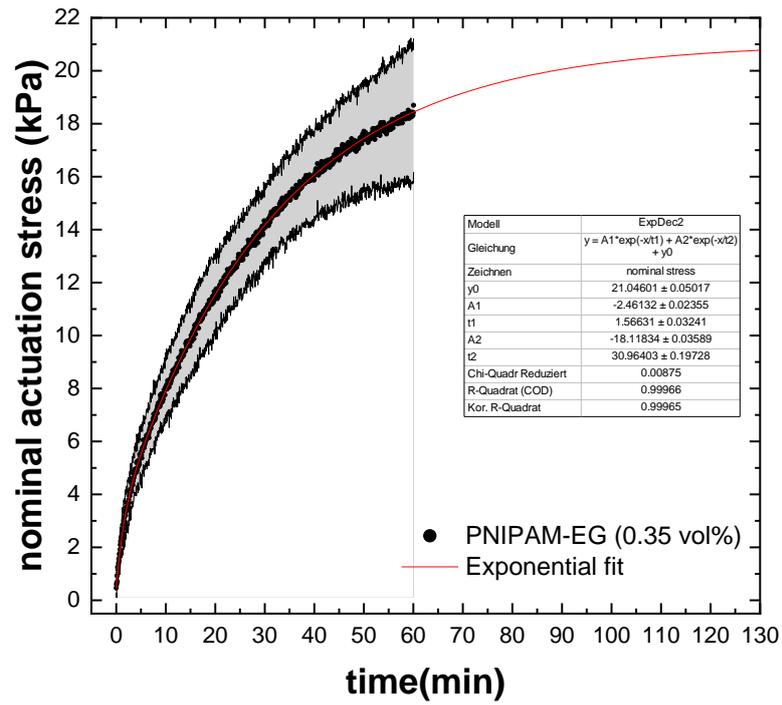

Figure S 6: Nominal actuation stress (black) of PNIPAM-EG (0.35 vol%) and exponential fit (red).

## Photothermal heating of PNIPAM-EG

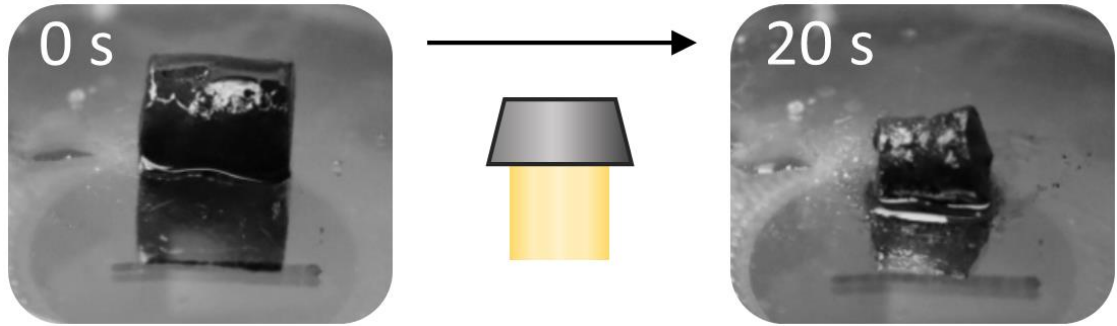

**Figure S7: Photothermal heating of PNIPAM-EG.** Illumination with white light for 20 s results in volume shrinkage.

## Deswelling and swelling properties of PNIPAM-structured and PNIPAM-EG with 16.1 % cross-linking

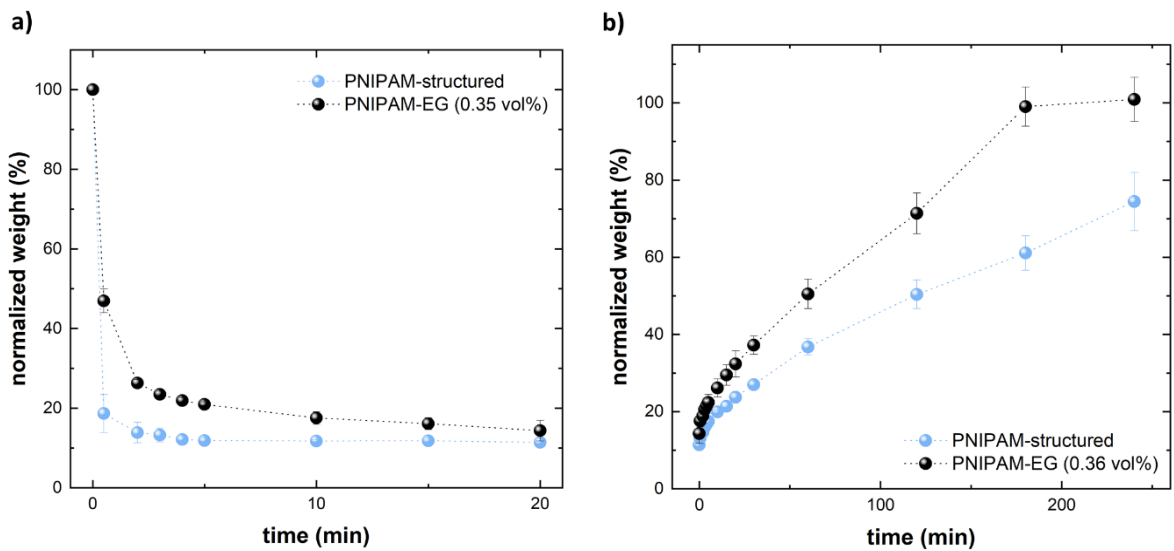

**Figure S8: Deswelling (a) and swelling (b) curves of PNIPAM-structured and PNIPAM-EG (0.35 vol%) with 16.1 % cross-linker.** Deswelling is slower for PNIPAM-EG compared to pure PNIPAM-structured hydrogels, while the samples reach almost the same total shrinkage. Swelling is enhanced for PNIPAM-EG samples, reaching the initial state after 3 h, whereas PNIPAM-structured only swell to about 60 % during this time.

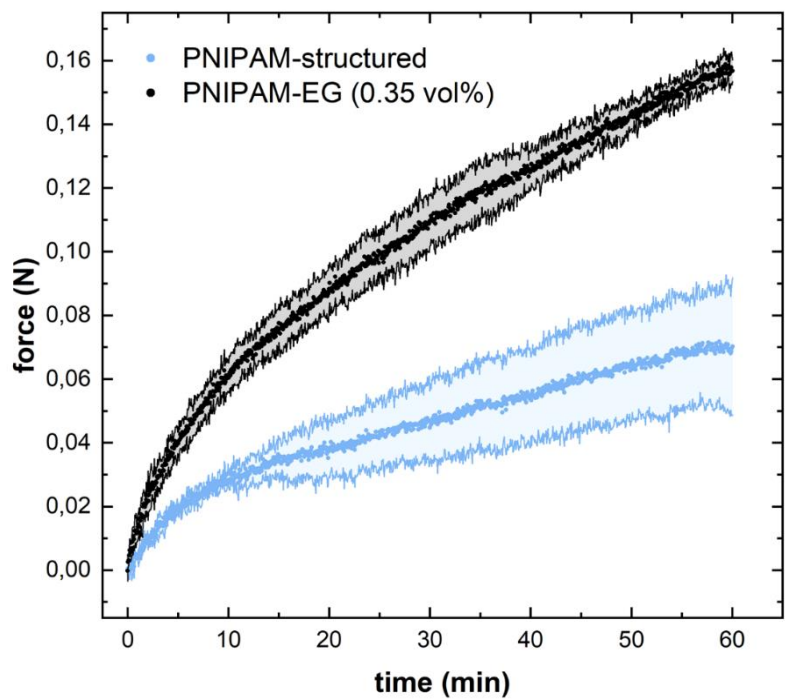

**Figure S9: Force exerted during swelling of PNIPAM-structured and PNIPAM-EG (0.35 vol%) with 16.1 % cross-linker.** The maximum exerted force for PNIPAM-EG (0.35 vol%) is 114 % higher compared to pure PNIPAM-structured.

**Joule heating of PNIPAM-EG**

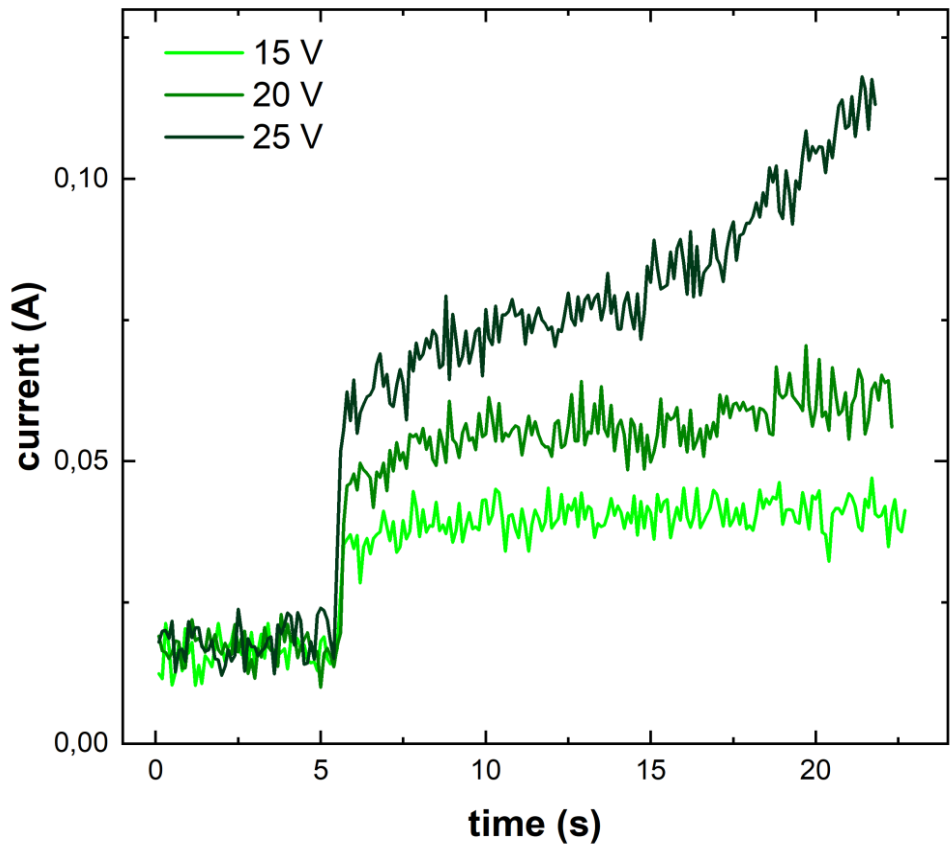

Figure S10: Current of PNIPAM-EG (0.35 vol % EG, 16.1 % cross-linker) for different applied voltages.

**SI Note 1 – Prediction of the Young's modulus of the graphene wall**

Based on experimental data the equivalent Young's modulus of the graphene tube is calculated to be 3.14 MPa by a direct rule of mixture. Following, the graphene wall Young's modulus is estimated. Note that assuming a bending dominated mechanism for the graphene tube wall would result in a scaling of the graphene Young's modulus with the square of the relative volumetric content of graphene within the tube. This can be estimated from the ratio of the volumetric contents of pore and graphene in the hydrogel, resulting in a prediction of the graphene wall Young's modulus around 3 GPa. This ratio also provides an estimation of the graphene thickness of around 32 nm. Note that an axial dominated deformative mechanism would result in the same scaling but with the power of 1 instead of 2, leading to an underestimated Young's modulus of the graphene wall of 0.1 GPa.

**SI Note 2 – Calculation of additional interfacial area**

The incorporation of an interconnected microtube network into the hydrogel matrix increases the significantly increases the hydrogel-water interfacial area.

A cylindrical bulk PNIPAM sample with a diameter of d = 6 mm and a height of h = 6 mm has a surface area $A_{bulk}$ of:

$$A_{bulk} = 2 * \pi * \left(\frac{d}{2}\right)^2 + 2 * \pi * \left(\frac{d}{2}\right) * h \tag{S1}$$

$$A_{bulk} = 1.696 \text{ cm}^2$$

The additional interfacial area created by the hollow microtubes can be determined from the surface area of the sacrificial tetrapodal ZnO template ($A_{template}$). For this, the surface area and the volume of a single ZnO tetrapod ($A_{tetrapod}$) are calculated, assuming that the single tetrapods are composed of 4 rods with a diameter $d_t$ = 2 µm and a length l of 25 µm.

$$A_{tetrapod} = 4 * \frac{2\pi * \frac{d_t}{2} * \left(\frac{d_t}{2} + l\right)}{10^8} \tag{S2}$$

$$V_{tetrapod} = 4 * \frac{\pi * \left(\frac{d_t}{2}\right)^2 * l}{10^{12}} \tag{S3}$$

Knowing the sample density ($\rho_{template}$ = 0.3 g cm$^{-3}$) and the density of ZnO ($\rho_{ZnO}$ = 5.61 g cm$^{-3}$), the number of tetrapods ($n_{tetrapods}$) in a cylindrical ZnO template with diameter of d = 6 mm and a height of h = 6 mm is calculated as follows:

$$n_{tetrapods} = \frac{V_{ZnO}}{V_{tetrapod}} = \frac{V_{sample} * \rho_{template}}{V_{tetrapod} * \rho_{ZnO}} \tag{S4}$$

With

$$V_{sample} = \frac{\pi * \left(\frac{d}{2}\right)^2 * h}{10} \tag{S5}$$

This results in an additional hydrogel-water interfacial area of:

$$A_{template} = n_{tetrapods} * A_{tetrapod} \tag{S6}$$

$$A_{template} = 188.697 \text{ cm}^2$$

The factor of additional hydrogel-water interfacial area is calculated to $\frac{A_{template}}{A_{bulk}} = 111.26$.

**Movie S1.**

Photothermal heating of PNIPAM-EG

**Movie S2.**

Joule heating of PNIPAM-EG

**Movie S3.**

Actuator designs